\documentclass[twocolumn,aps,prl,showpacs,superscriptaddress,10pt] {revtex4-1}

	\usepackage{amsmath}%,amssymb} 
	\usepackage{makeidx}
	\usepackage{amsfonts}
	\usepackage[ansinew]{inputenc}
	\usepackage[usenames,dvipsnames]{pstricks}
	\usepackage{subfigure}
	\usepackage{epsfig}
	\usepackage{pst-grad} % For gradients
	\usepackage{pst-plot} % For axes
	\usepackage[hyperindex]{hyperref}
	\usepackage{color}
\usepackage{wasysym}
	
\makeindex

\begin{document}

\title{Dephasing in strongly anisotropic black phosphorus}

\author{N. Hemsworth}
\author{V. Tayari}
\affiliation{Department of Electrical and Computer Engineering, McGill University, Montr\'eal, Qu\'ebec, H3A 2A7, Canada}
\author{F. Telesio}
\author{S. Xiang}
\author{S. Roddaro}
\affiliation{NEST, Istituto Nanoscienze-CNR and Scuola Normale Superiore, Pisa, Italy}
\author{M. Caporali}
\author{A. Ienco}
\author{M. Serrano-Ruiz}
\author{M. Peruzzini}
\affiliation{Istituto Chimica dei Composti OrganoMetallici-CNR, Sesto Fiorentino, Italy}
\author{G. Gervais$^{*}$}
\affiliation{Physics Department, McGill University, Montr\'eal, Qu\'ebec, H3A 2T8, Canada}
\affiliation{$^{*}$corresponding author: gervais@physics.mcgill.ca}
\author{T. Szkopek}
\affiliation{Department of Electrical and Computer Engineering, McGill University, Montr\'eal, Qu\'ebec, H3A 2A7, Canada}
\author{S. Heun}
\affiliation{NEST, Istituto Nanoscienze-CNR and Scuola Normale Superiore, Pisa, Italy}

%\date{today}

\begin{abstract}
Weak localization was observed in a black phosphorus (bP) field-effect transistor 65 $nm$ thick. The weak localization behaviour was found to be in excellent agreement with the Hikami-Larkin-Nagaoka model for fields up to 1~T, from which characteristic scattering lengths could be inferred. The dephasing length $L_{\phi}$ was found to increase linearly with increasing hole density attaining a maximum value of $55~\mathrm{nm}$ at a hole density of approximately $10^{13} \mathrm{cm^{-2}}$ inferred from the Hall effect. The temperature dependence of $L_{\phi}$ was also investigated and above 1~K, it was found to decrease weaker than the $L_{\phi} \propto T^{-\frac{1}{2}}$ dependence characteristic of electron-electron scattering in the presence of elastic scattering in two dimensions. Rather, the observed power law was found to be close to that observed previously in other quasi-one-dimensional systems such as metallic nanowires and carbon nanotubes. We attribute our result to the crystal structure of bP which host a `puckered' honeycomb lattice forming a strongly anisotropic medium for localization.   \end{abstract}

%\pacs{65.80.Ck}

\maketitle

% Introduction + Background bP

The recent surge of interest in the field of 2D  atomic crystals has led to a number of important advances in our understanding of solid-state physics in two dimensions.  These include transition metal dichalcogenides, topological insulators such as bismuth selenide, and phosphorene. For the latter,  its underlying crystal structure  inherited from black phosphorus (bP) is a `puckered'  honeycomb lattice hosting one-dimensional chains of atoms displaced out of the atomic plane of the honeycomb lattice, and running along the zig zag axis. Few-layer black phosphorus has been the subject of recent investigations wherein strong anisotropy was observed in electronic transport, optical absorption, thermal conductivity and angle-resolved photoemission measurements \cite{Xia2014,Wang2015,Luo2015,Kim2015}. Here, we have studied the weak localization in a bP field-effect transistor and we found that the phase coherence length decays with temperature at a slower rate than that expected for a two-dimensional material. This more robust coherence is reminiscent to that previously reported in quasi-one-dimensional systems such as nanotubes\cite{Appenzeller2001} and metallic nanowires\cite{Natelson2001}. Our experimental study of weak localization in bP is amongst very few investigations of weak localization in strongly anisotropic media.

Unlike semi-metallic graphene, bP is a direct gap semiconductor \cite{Keyes1953,Morita1986}. The bandgap is 0.3~eV in bulk  and increases by quantum confinement to 1-2~eV in the monolayer limit \cite{Du2010,Das2014,Liang2014}, ideal for applications in electronics and optoelectronics \cite{Castellanos2015}.  Black phosphorus field effect transistors (FETs) have been demonstrated by various groups \cite{Li2014,Xia2014,Liu2014,Koenig2014,Gomez2014,Wood2014}, with hole field effect mobilities reaching up to $1350~\mathrm{cm^2/Vs}$ and $10^5$ current modulation at room temperature \cite{Chen2015}. In the presence of a large magnetic field normal to the bP atomic layers, Shubnikov-de Haas oscillations have been observed in bP FETs \cite{Li2015,Tayari2015,Chen2015,Gillgren2015}. \\

\begin{figure}[tbp]
	\begin{center}
		\includegraphics[width=.9\linewidth]{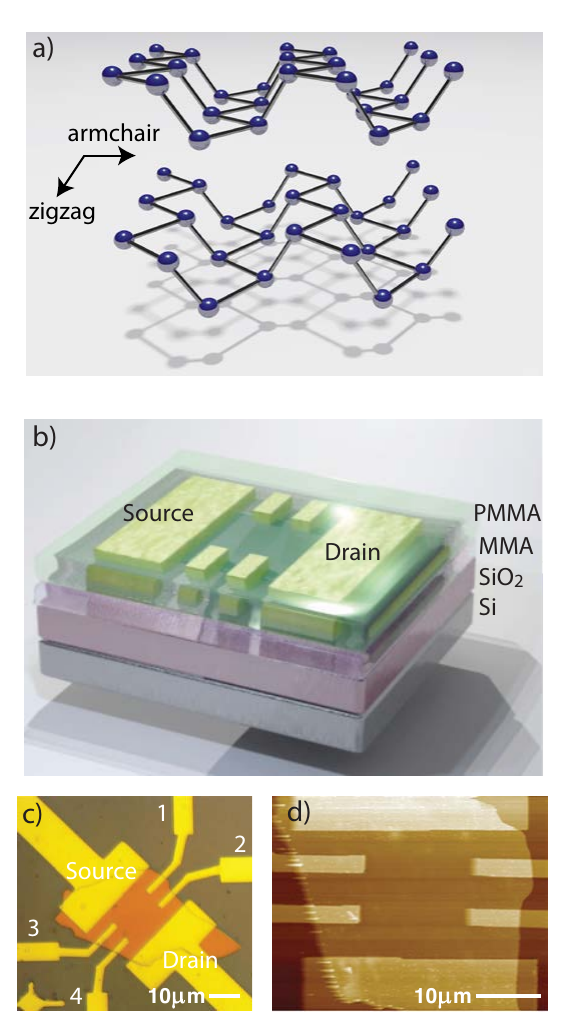} 
		\caption{{\bf Black phosphorus field-effect transistor}. a) Cartoon of the puckered honeycomb crystal structure of bP.  b) Schematic of the back-gated bP FET in a Hall bar geometry. c) Optical reflection image of the bP FET. The longitudinal resistance $R_{xx}$ was measured with voltage probes 1 and 2, and the Hall resistance $R_{xy}$ was measured with voltage probes 1 and 3. d) AFM image of the bP FET. The bP thickness was measured to be $65\pm2$~nm. }
		\label{fig:Schematic}
	\end{center}
\end{figure}

Magnetotransport in the presence of a weak magnetic field normal to the bP atomic layers is less well studied. In a disordered 2-D system, coherent backscattering of charge carriers gives rise to a peak in the magnetoresistance known as weak localization (WL),  see \cite{Bergmann1984} for a review. Previous work on weak localization in 2D crystals has focused primarily on graphene \cite{Morozov2006,Wu2007,Tikhonenko2008} and has also been observed recently in molybdenum disulfide, revealing a phase coherence length $L_{\phi}$ of $\sim 50$ $\mathrm{nm}$ at $400$~$\mathrm{mK}$ that decays with increasing temperature as $T^{-\alpha}\simeq T^{-\frac{1}{2}}$ \cite{Neal2013}, characteristic of electron-electron interactions in 2D. Dimensionality plays an important role in weak localization. Previous work in metal and semiconducting mesoscopic structures has shown the dephasing length exponent $\alpha$ to decrease from $1/2$ to a value close to $1/3$ as the system geometry was reduced towards the one dimensional limit \cite{Natelson2001}, in agreement with theory of weak localization \cite{Altshuler1982}. Here,  we have observed weak localization in a  65~nm thick bP FET whose dephasing length exponent is found to be suppressed, and closer to a value of 1/3. We attribute this weak localization behaviour to the anisotropic nature of the puckered bP atomic crystal structure whose holes dispersion is distinct along the hard (along) and easy (against) axis of the puckers, see Fig.\ref{fig:Schematic}.\\

%Weak localization has been observed in bP, with a phase coherence length of $\sim$$100\mathrm{nm}$ at $350~\mathrm{mK}$ that also decayed with temperature as $T^{-1/2}$ for $T>6~\mathrm{K}$ \cite{Du2016}.

% In this paper.... (describe the device + fabrication)

% In this paper.... (describe the device + fabrication)

% Measurement Methodology

\begin{figure}[tbp]
	\begin{center}
		\includegraphics[width=.9\linewidth]{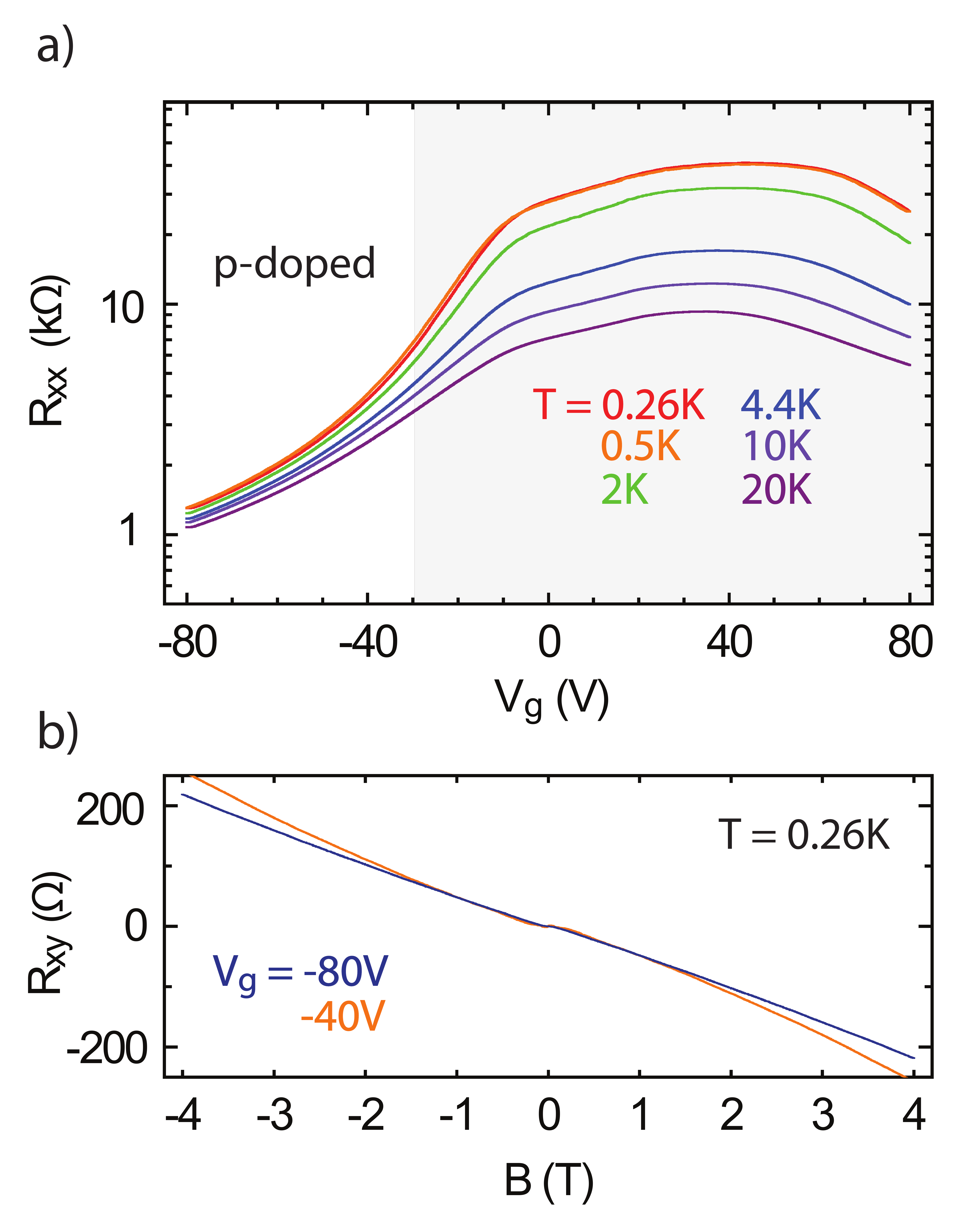} 
		\caption{{\bf Transport measurements versus backgate voltage and temperature}.  a) The measured longitudinal resistance $R_{xx}$ versus gate voltage $V_g$, at temperatures $T=0.26~\mathrm{K}$ to $20~\mathrm{K}$. The region of strong p-doping is identified.   b) The Hall resistance $R_{xy}$ versus magnetic field $B$ at $T=0.26~\mathrm{K}$, with the component symmetric in $B$ removed, as measured at $V_g = -80~\mathrm{V}$ and $-40~\mathrm{V}$.}
		\label{fig:Transport}
	\end{center}
\end{figure}

% Carrier density
% WL in graphene and other 2D? short paragraph maybe? Graphene, bilayer graphene, MoS2. 
The design, as well as  photograph and atomic force microscope images of the bP FET are shown in Fig.\ref{fig:Schematic}. The longitudinal resistance $R_{xx}$ versus back gate voltage $V_g$ is plotted in Fig.~\ref{fig:Transport}(a), with sample temperature $T$ as a parameter. The sample is highly resistive for $V_g > -30~\mathrm{V}$, and exhibits clear p-type conduction for $V_g < -30~\mathrm{V}$. The Hall resistance $R_{xy}$ versus magnetic field $B$ is plotted in Fig.~\ref{fig:Transport}(b) at $T = 0.26~\mathrm{K}$ and at gate voltage $V_g = -40~\mathrm{V}$ and $-80~\mathrm{V}$, with the component symmetric in $B$ removed. Hole carrier density $p$ inferred from the Hall resistance is found to depend linearly on gate voltage $V_g$ over the range of p-type conduction $V_g < -30~\mathrm{V}$ with a density of approximately $10^{13}/$ cm$^{2}$ at $V_g = -30$~V. The field-effect hole mobility, $\mu_{FE} = (L/W) \cdot \partial G_{xx}/\partial ( C_g V_g) $ was found to reach a peak value of $300 \mathrm{cm^2/Vs}$ at a gate voltage $V_g = -70~\mathrm{V}$, with a negligible dependence upon temperature over the measured range $0.26~\mathrm{K} < T < 20~\mathrm{K}$. \\

% WL & Fitting

\begin{figure}[tbp]
	\begin{center}
		\includegraphics[width=.95\linewidth]{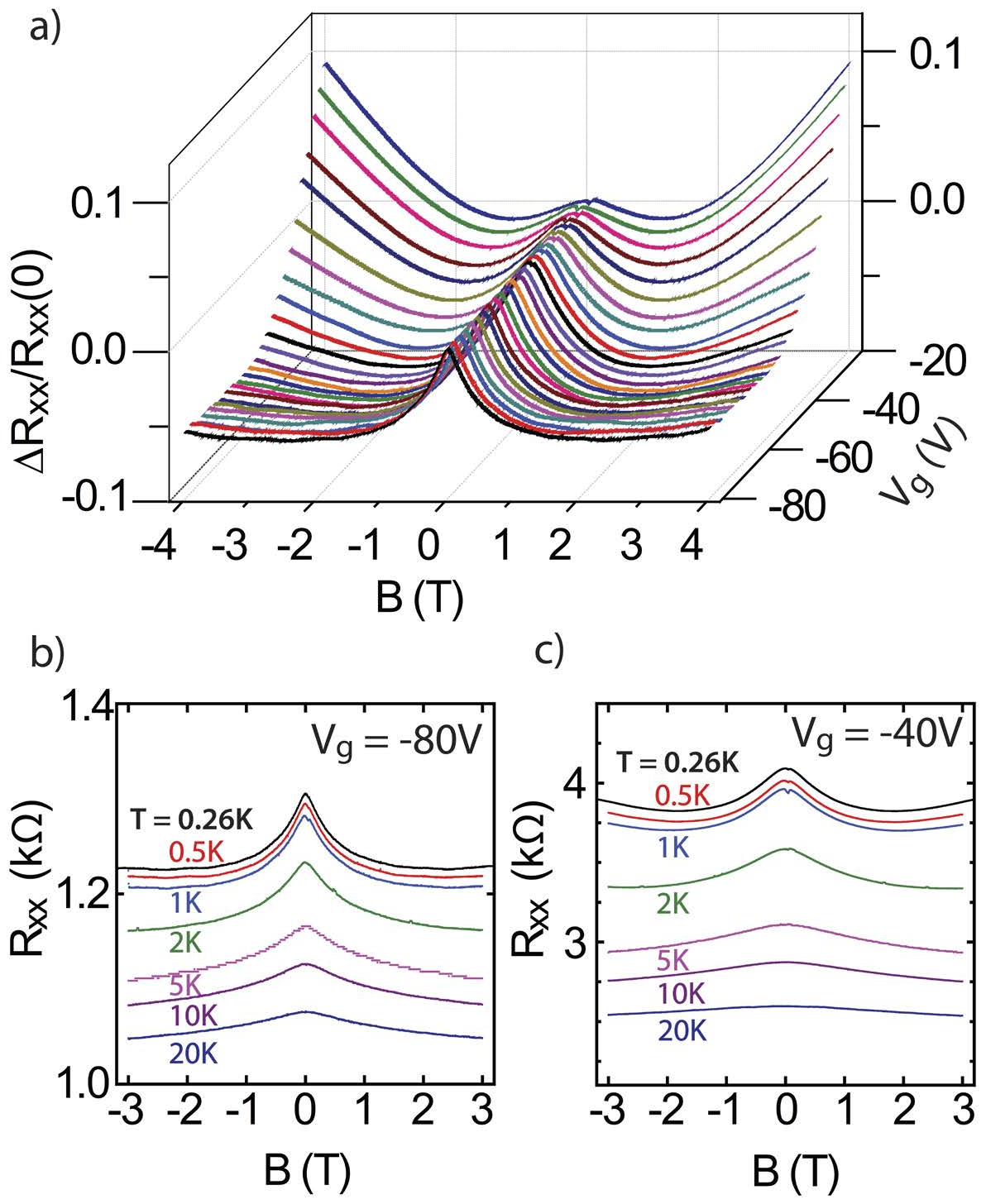} 
		\caption{{\bf Weak localization measurements}. a) A weak localization peak is observed in a plot of the normalized longitudinal resistance $(R_{xx}(B)-R_{xx}(0))/R_{xx}(0)$ versus magnetic field $B$ and gate voltage $V_{g}$ at $T=0.26~\mathrm{K}$. The temperature dependence of the longitudinal resistance $R_{xx}$ versus magnetic field $B$ at b) $V_g=-80~\mathrm{V}$ and at c) $V_g=-40~\mathrm{V}$.}
		\label{fig:Rxx}
	\end{center}
\end{figure}

The measured weak localization peak in longitudinal resistance is shown in the plot of $\Delta R_{xx} / R_{xx}(0) = ( R_{xx}(B)-R_{xx}(0) ) / R_{xx}(0)$ versus magnetic field $B$ and gate voltage $V_g$ in Fig.~\ref{fig:Rxx}(a). The amplitude of the WL peak increases with increasing hole density attaining a maximum value at the highest negative gate voltage used in this experiment, -80V. The temperature dependence of the WL peak at $V_g = -80~\mathrm{V}$ and $V_g = -40~\mathrm{V}$ is plotted in Fig.~\ref{fig:Rxx}(b) and Fig.~\ref{fig:Rxx}(c), respectively. The WL  correction to the resistance decreases with increasing temperature eventually disappearing at temperatures above 20~K, as expected. \\

The Hikami-Larkin-Nagaoka (HLN) theory \cite{Hikami1980,Maekawa1981,Bergmann1984} gives a quantitative prediction for the WL correction to the sheet conductance,
\begin{eqnarray}
\Delta \sigma & = & -\frac{e^{2}}{2\pi^2\hbar}\left(\Psi\left(\frac{1}{2} + \frac{B_1}{B}\right)-\Psi\left(\frac{1}{2} + \frac{B_2}{B}\right) \right. \nonumber \\
& & \left. + \frac{1}{2}\Psi\left(\frac{1}{2} + \frac{B_3}{B}\right) - \frac{1}{2}\Psi\left(\frac{1}{2} + \frac{B_2}{B}\right)\right),
\end{eqnarray}
where $\Psi$ is the digamma function. The field parameters in the above expression are given by:
\begin{eqnarray}
B_{1} & = & B_{0} + B_{so} + B_{s}\\
B_{2} & = & \frac{4}{3}B_{so} + \frac{2}{3}B_{s} + B_{\phi}\\
B_{3} & = & 2B_{s} + B_{\phi}
\end{eqnarray}
where $B_{0}$, $B_{so}$, $B_{s}$, and $B_{\phi}$ are the characteristic fields associated with elastic scattering, spin-orbit scattering, magnetic scattering, and inelastic scattering (dephasing), respectively.\\

\begin{figure}[tbp]
	\begin{center}
		\includegraphics[width=1.05\linewidth]{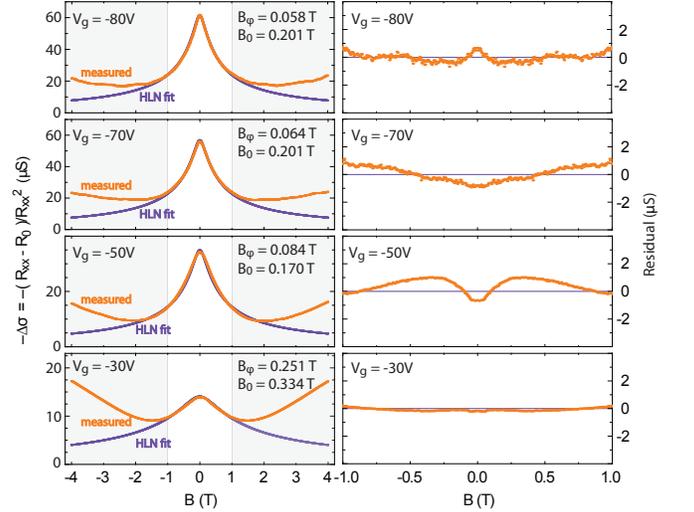} 
		\caption{{\bf HLN weak localization fits}. The measured longitudinal conductance correction $-\Delta \sigma$ versus magnetic field $B$ and a best fit to a Hikami-Larkin-Nagaoka theory over the field range $-1~\mathrm{T}<B<1~\mathrm{T}$, where weak localization dominates, is shown for various gate voltages $V_g$ at $T=0.26~\mathrm{K}$. The extracted characteristic fields $B_{0}$ for elastic and $B_{\phi}$ for inelastic (dephasing) scattering are indicated.}
		\label{fig:RxxFit}
	\end{center}
\end{figure}

The measured WL peak (normalized) resistances were inverted into conductivity by the usual tensor relation $\sigma_{xx}=\rho_{xx}/(\rho_{xx}^2+\rho_{xy}^2)$, with negligible contributions from the (transverse) Hall resistivity (see \ref{fig:Transport}(b)). To account for the background resistivity, the  relation $\Delta \sigma = - (L/W) \cdot ( R_{xx}(B) - R_{0} )/R_{xx}(B)^2$ was used,  where  $R_0$ is the classical Drude resistance in the absence of a WL correction. The determination of the  $R_{0}$ is not trivial, and was thus left as a fit parameter, however it was verified that its trend in temperature followed that of the measured transport mobility of the bP.  This WL correction to conductivity was numerically fitted to the HLN model under the approximation of negligible spin-orbit coupling, $B_{so} = 0$, and negligible magnetic impurity scattering, $B_{s}=0$. Both approximations are appropriate for our bP crystals \cite{ICP}. Measurements and numerical fits of the WL contribution to conductivity $\Delta \sigma$ are plotted in Fig.~\ref{fig:RxxFit} versus backgate voltage at temperature $T=0.26$K. The fit was performed over the magnetic field range $-1~\mathrm{T}<  B < +1\mathrm{T}$, where the WL feature dominates the magnetoresistance. The fit quality was evaluated over the  same $B$-range, with an $R^2$ coefficient of determination of at minimum 0.99. The residuals are also shown in the right panels of Fig.~\ref{fig:RxxFit}. \\

\begin{figure}[tbp]
	\begin{center}
		\includegraphics[width=.9\linewidth]{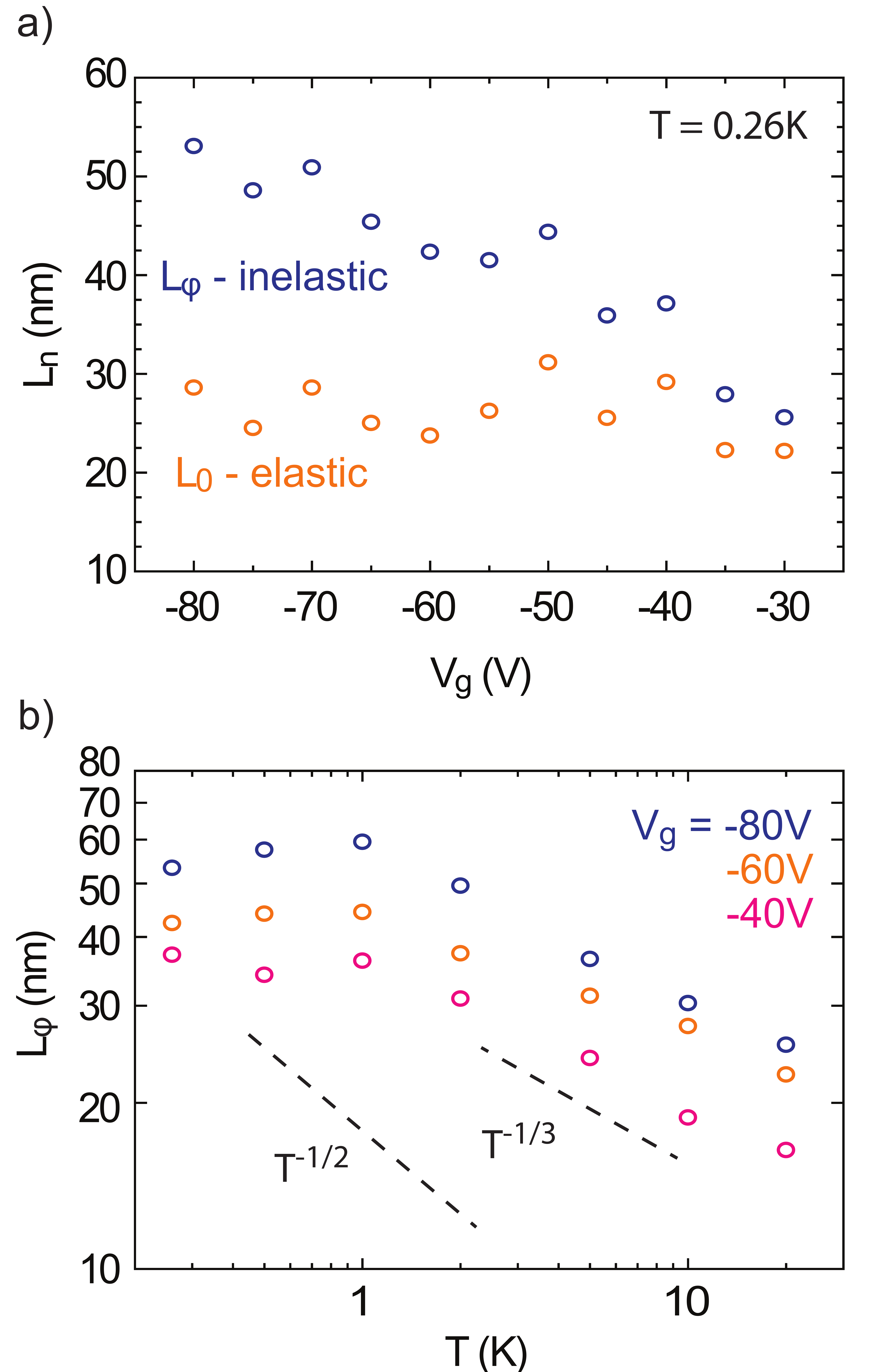} 
		\caption{{\bf Scattering lengths and power laws}. a) The inelastic scattering length $L_{\phi}$ and elastic scattering length $L_0$ versus gate voltage $V_g$, at temperature $T=0.26~\mathrm{K}$. b) The inelastic scattering length $L_{\phi}$ versus temperature $T$ at various gate voltages $V_g$. The $T^{-1/2}$ temperature dependence associated with electron-electron scattering in the 2D diffusive limit is shown for comparison, as well as a $T^{-1/3}$ power law.}
		\label{fig:ScatteringTimes}
	\end{center}
\end{figure}

The extracted characteristic fields $B_{0,\phi}$ are related to the scattering lengths $L_{0,\phi}$ by considering the phase shift of diffusing charge carriers under the influence of a magnetic field, $B L^2 = \hbar /4e$. From these fields, we can deduce elastic as well as inelastic (dephasing) characteristic  lengths, $L_0$ and $L_{\phi}$ respectively. These lengths are plotted versus gate voltage $V_g$ in Fig.~\ref{fig:ScatteringTimes}(a) at $T = 0.26~\mathrm{K}$. In particular, the dephasing length $L_{\phi}$ clearly increases with increasing hole density, reaching a maximum of 55~nm, whereas $L_{0}$ remains  nearly independent over a broad range of carrier density. We have also verified that $L_{0}$ scales linearly with  the transport mobility, as expected for an elastic scattering process occurring in the presence of both a phonon bath and disorder.  \\

For weak localization occurring in typical disordered two-dimensional systems, the dephasing length is related to the inelastic scattering time via $L_\phi^2 = D \tau_\phi$, where $D$ is the elastic diffusion coefficient. Electron-electron scattering in the absence of elastic scattering is expected to give a scattering rate $1/\tau_{\phi} \propto T^2$, and hence $L_{\phi} \propto T^{-1}$ \cite{Tikhonenko2009}. In the diffusive transport regime, appropriate here since $L_o < L_{\phi}$, the electron-electron scattering rate is expected to follow $1/\tau_{\phi} \propto T$ \cite{Altshuler1980,Abrahams1982,Altshuler1982}, and hence $L_{\phi} \propto T^{-1/2}$. This characteristic inelastic scattering rate has been observed in graphene \cite{Morozov2006,Wu2007,Tikhonenko2008,Tikhonenko2009} and MoS$_2$ \cite{Neal2013}, however our weak localization measurements in a bP thin film clearly de not follow this trend. The temperature dependence of $L_{\phi}$ is shown in Fig.~\ref{fig:ScatteringTimes}(b) in a log-log plot at fixed gate voltages $V_g = -40, -60, -80~\mathrm{V}$.  The experimental error in $L_{\phi}$ was determined by way of visual inspection of the quality of the fit when varying one parameter and keeping all others constant, and are contained within the size of the symbols. The saturation of $L_{\phi}$ at temperatures below 1~K is most likely due to impurities, as previously observed in  a variety of metallic and semiconducting 2D systems \cite{Lin2002b}. However, above 1~K the temperature dependence of $L_{\phi}$ does \underline{not} follow the $T^{-\frac{1}{2}}$ (dashed line)  behaviour expected  {\it a priori}   from electron-electron scattering in the presence of elastic scattering in 2D. A weighted fit of the inelastic scattering length versus temperature rather show a suppressed dephasing length exponent (and corresponding 95\% confidence interval withing bracket) $\alpha$ of  $0.29~ [0.26-0.32] , 0.22~ [0.19-0.25], 0.27~[0.23-0.31]$, with $R^2$ coefficients all greater than 0.99 at  gate voltages $V_g = -80, -60, -40~\mathrm{V}$, respectively. \\

\begin{figure}[tbp]
	\begin{center}
		\includegraphics[width=.9\linewidth]{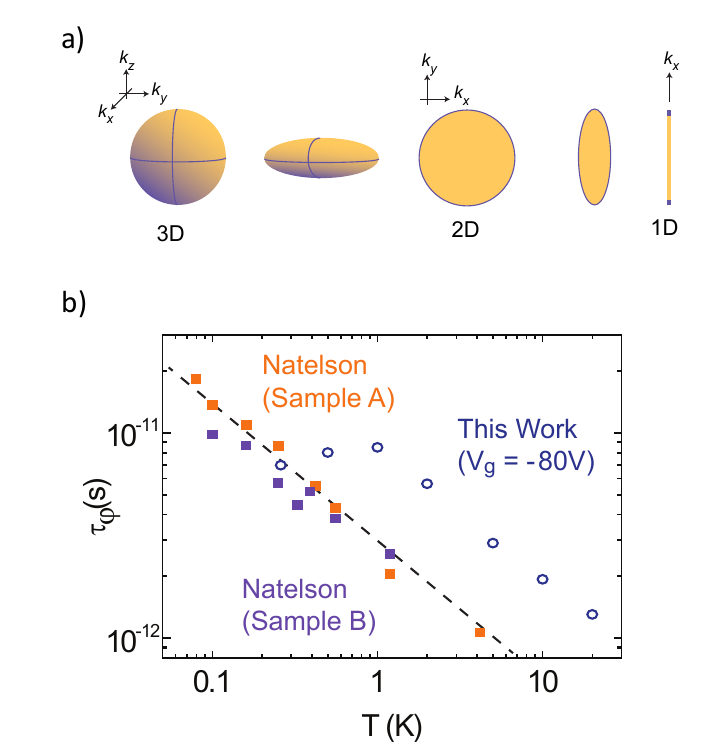} 
		\caption{{\bf Comparison with quasi-one-dimensional metallic wire}. a) Schematuic depicting the Fermi surface evolution from 3D to 1D. b) The data in bP at gate voltage $V_{g}=-80V$ converted into dephasing time $\tau_{\phi}$  are compared  with those of Natelson {\it et al.}\cite{Natelson2001} measured in metallic nanowires in the quasi-1D regime. The dotted lines shows a dephasing time decaying as $T^{-\frac{2}{3}}$, {\it i.e.} corresponding to a power law exponent $\alpha=\frac{1}{3}$ for $L_{\phi}$.}
		\label{fig:tauphi}
	\end{center}
\end{figure}

Our WL mesurements reveal the presence of an inelastic scattering mechanism that is characterized by a decay in dephasing length weaker than $T^{-1/2}$. This is in contrast with previous measurements in a few-layer bP films \cite{Du2016} where a power law was also deduced, albeit with saturation of $L_{\phi}$ at higher temperatures than observed here (5~K as opposed to 1~K in our case). The analysis of \cite{Du2016} included a fitting parameter for valley degeneracy in bP. This parameter was found to differ from unity (no valley degeneracy) and to vary from $\sim$1.2 to 0.3 over the range of temperatures where the power law exponent for $L_{\phi}$ was extracted. The band structure of bP has been shown by photoemission studies \cite{Li2014}, magnetotransport\cite{Tayari2015}, and density functional theory\cite{Qiao2014} to be absent of any valley degeneracies, calling into question the robustness of the  $T^{-1/2}$ scaling reported in Ref.\cite{Du2016}.   \\

A weaker loss of the dephasing length with temperature, as compared to the 2D case, has previously been observed in the presence of a strong 1D confinement potential, {\it e.g.} in carbon nanotubes \cite{Appenzeller2001}, and $L_{\phi}$ was shown to be well described by the theoretical prediction of a $T^{-\frac{1}{3}}$ power law. Remarkably, the transition from the 2D limit to the quasi 1D limit was observed in the systematic study undertaken by Natelson {\it et al.}\cite{Natelson2001} whereby the dephasing time in metallic nanowires was measured as a function of the wire width down to 5~nm. In this regime of width and low temperatures, several heuristic length scales can be estimated placing the wires well into a quasi-1D regime. In this study a lower saturation temperature was observed,  however  the dephasing time $\tau_{\phi}$ was determined to follow closely  a $T^{-\frac{2}{3}}$ power law, corresponding to a dephasing length exponent  $\alpha=\frac{1}{3}$. Fig.\ref{fig:tauphi} shows a comparison between Natelson's dephasing time and the dephasing time of our bP FET at $V_g = -80$~V. The expected $T^{-\frac{2}{3}}$ power law for $\tau_{\phi}$ in the quasi-1D limit is also shown as a dotted line.  \\

The robustness against dephasing in bP, when compared to the ideal 2D case in the presence of electron-electron scattering, is  {\it a priori} unexpected. Years of previous work have shown weak localization (and hence dephasing) to be sensitive to system's dimensionality. To our knowledge, there are two known scenarios for which a power-law exponent weaker than from $\alpha = \frac{1}{2}$  can occur. These are {\it i)} strong confinement due to a 1D potential, as mentioned above and {\it ii)} dephasing by edge domain granularity \cite{Beliayev2011}. The crystalline order in our bP FET was carefully investigated by way of both unpolarized and polarized micro-Raman measurements and these data strongly suggest the bP flake to be an homogeneous single crystal (see supplementary information) rendering the granularity scenario unlikely. It is therefore more likely that the anisotropic nature of bP arising from puckers in the atomic plane  which give rise to a an ellipsoid Fermi surface with heavy hole mass in the zig zag direction parallel with the puckers, and light hole mass in the armchair direction perpendicular to the puckers. Making use of calculated values for the effective masses as reported in Ref.\cite{Qiao2014}, we estimate the corresponding Fermi wavelengths along and against the puckers to be $\lambda_{F,zz}=5$ and $\lambda_{F,ac}=12$ nm, respectively, at least one order of magnitude larger than any atomic lengthscales. In the basal direction, the presence of an electric field from the backgate has been shown to create a two-dimensional hole gas  with a thickness $\delta_{2deg}$ of $\approx 3$ nm \cite{Li2015, Tayari2015}, a scale that is much less than the thermal length $L_T=\sqrt{\hbar D/k_BT} \sim 10-60 $ nm  over all temperaturea investigated in our experiment. These lengthscales place the bP FET into a strongly anisotropic regime for which weak, and even strong localization, has not yet been well studied.\\

%In summary, weak localization was observed in a 65~nm thick black phosphorus field-effect transistor and studied over a broad range of gate voltages (hole density) and temperatures. 

The measured weak localization demonstrates a more robust coherence observed as a function of temperature as previously observed in quasi-one-dimensional  systems such as nanotubes and metallic nanowires. Since it is highly sensitive to  dimensionality, we attributed the more robust character of $L_{\phi}$ in bP to the highly anisotropic nature of the puckered honeycomb crystal structure. In the future,   it will be of  great interest to understand exactly how dephasing is affected by strong anisotropy in post-graphene 2D atomic crystals, a subject that is currently nearly free of any theoretical knowledge.\\

We thank Vincenzo Piazza for his help with the polarized Raman measurements. The authors thanks NSERC (Canada), Cifar (Canada), the Canada Research Chairs Program, Hydro-Qu\'ebec, FRQNT (Qu\'ebec) and the Institut de l'Energie Trottier for support. We also thank the European Research Council (ERC) under the European Union's Horizon 2020 research and innovation program (Grant Agreement No. 670173) for funding the project PHOSFUN `Phosphorene functionalization: a new platform for advanced multifunctional materials' through an ERC Advanced Grant. We acknowledge funding from the Italian Ministry of Foreign Affairs, Direzione Generale per la Promozione del Sistema Paese (agreements on scientific collaboration with Canada (Quebec) and Poland). Financial support from the CNR in the framework of the agreements on scientific collaboration between CNR and NRF (Korea), CNRS (France), and RFBR (Russia) is acknowledged. Furthermore, funding from the European Union Seventh Framework Programme under Grant Agreement No. 604391 Graphene Flagship is acknowledged. G.G. acknowledges two short-term mobility grants from CNR.\\

{\bf Special Note}: upon the completion of this work and posting on the arXiv, we have become aware of similar results obtained by the UC Riverside group (Lau), {\it arXiv}:1608.00323.

\newpage

%\subsection{HLN Fits at 10K}

{\bf \large Supplementary material for ``Dephasing in strongly anisotropic black phosphorus''}

%\affiliation{$^{1}$ Department of Physics, McGill University,Montreal, H3A 2T8, CANADA}

%\affiliation{$^{2}$ Department of Electrical Engineering, Princeton University, Princeton NJ 08544 USA}

\vspace*{5mm}

{\bf Methods} \\

Our bP FETs were prepared by mechanical exfoliation of bulk bP crystals onto a degenerately doped Si wafer with a $300~\mathrm{nm}$ SiO$_2$ layer prepared by dry thermal oxidation.  The bP crystals were prepared by heating commercially available red phosphorus ($>99.99\%$) in a muffle oven, together with a tin ($>99.999\%$), gold ($>99.99\%$),  and a catalytic amount of SnI$_4$, following a published procedure \cite{Nilges2008}. The solids were loaded in a quartz tube, which was then evacuated by a pumping procedure where the vacuum was backfilled by $N_2$ gas several times and then  the tube sealed under vacuum. Then, it was heated to to $406^\circ$C (at a rate of $4.2^\circ$C/min),  kept 2h at this temperature and then heated up to 650ºC ($2.2^\circ$C/min). The sample stayed for three days at this temperature in the oven. Afterwards, a slow cooling rate was chosen ($0.1^\circ$C/min) to afford the formation of crystals of bP (typical size: 2 mm $\times$ 3 mm). A photograph of a typical crystal is shown in the supplementary information. Exfoliation was performed in a nitrogen glove box to suppress photo-oxidation, and the SiO$_2$ surface was treated with a hexamethyldisilazane (HMDS) layer to suppress charge transfer doping. Standard electron beam lithography was used to define 5~nm Ti/ 80~nm Au metal electrodes in a Hall bar geometry. The samples were made environmentally stable by encapsulation with $300~\mathrm{nm}$ of methyl methacrylate (MMA) and $200~\mathrm{nm}$ of polymethyl methacrylate (PMMA). The thickness of the bP was determined to be $65\pm2$~nm by AFM. Importantly, the hole accumulation layer induced in a bP FET at cryogenic temperatures is estimated to be $<3~\mathrm{nm}$ thick, corresponding to a hole gas occupying 5-6 bP layers \cite{Tayari2016}. Charge transport measurements were performed using standard ac lock-in techniques in a $^3$He cryostat with sample temperature ranging from $T=0.26$~K to $T=20$~K. A $100~\mathrm{nA}$ ac current at frequency $f=17$ Hz was applied through the source and drain terminals of the bP FET and both the longitudinal and transverse (Hall) voltages were measured. A dc gate voltage $V_{g}$ was applied to tune the carrier density through the back gate capacitance $C_g = 11.5~\mathrm{nF/cm^2}$.

\vspace*{5mm}

\begin{figure}[t]
	\begin{center}
		\includegraphics[width=.8\linewidth]{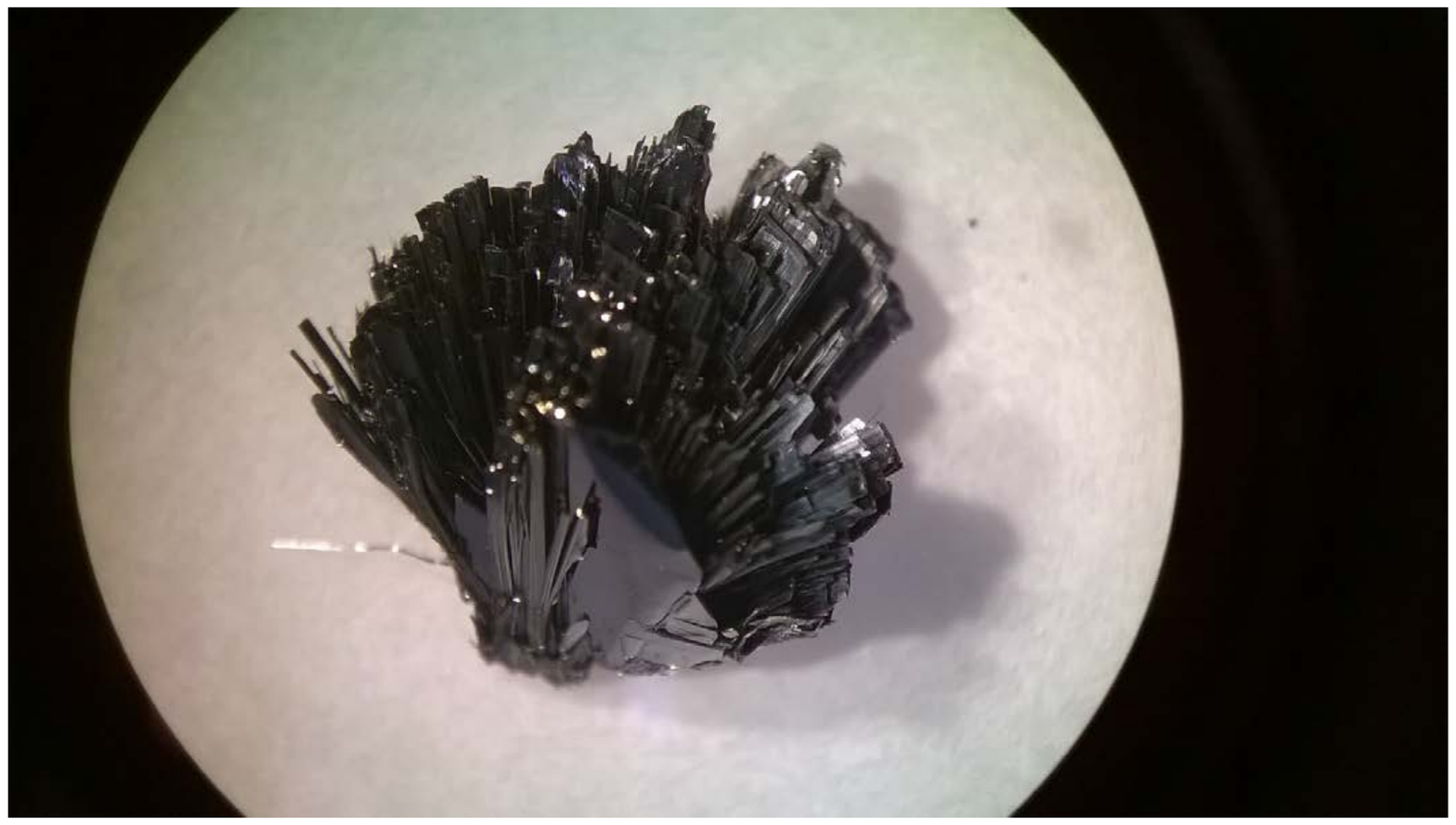}
		\caption{{\bf Photograph of the synthesized black phosphorus}. The crystals were synthesized per the technique mentioned in the methods section. }
		\label{fig:bP}
	\end{center}
\end{figure}

\begin{figure}[t]
	\begin{center}
		\includegraphics[width=.8\linewidth]{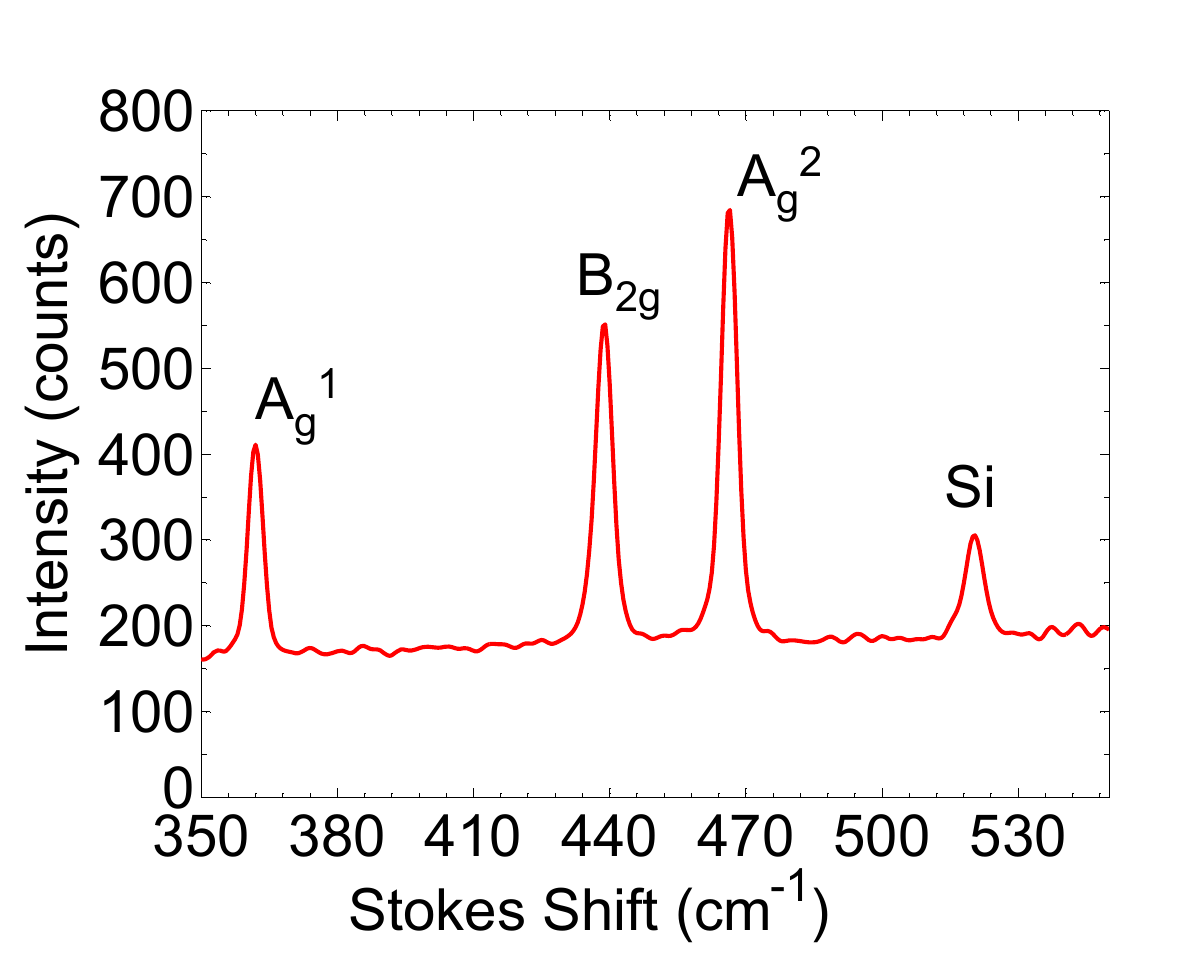} 
		\caption{{\bf Unpolarized Raman spectrum}. The Raman Stokes shift of encapsulated bP FET was measured with a 532 nm laser pump and 5 mW of incident power to enable acquisition through the semi-transparent gold layer. A silicon substrate peak at 520 cm$^{-1}$ is observed, along with the A$_g^1$, B$_g^2$ and A$_g^2$ peaks characteristic of bP at 362 cm$^{-1}$, 439 cm$^{-1}$, and 466.5 cm$^{-1}$, respectively. }
		\label{fig:UnpolarizedRaman}
	\end{center}
\end{figure}

{\bf Unpolarized Raman Characterization}\\

The Raman Stokes spectrum of the bP sample was measured after encapsulation by a 6 nm thick layer of gold for long term storage. Spectra were acquired with a micro-Raman spectrometer with a 532 nm laser pump and 5 mW of incident power to enable acquisition through the semi-transparent gold layer. A representative spectrum is shown in Fig.\ref{fig:UnpolarizedRaman}. A silicon substrate peak at 520 cm$^{-1}$ is observed, along with the A$_g^1$, B$_{2g}$ and A$_g^2$ peaks characteristic of bP at 362 cm$^{-1}$, 439 cm$^{-1}$, and 466.5 cm$^{-1}$, respectively. A red-shift of 3-7 cm$^{-1}$ compared to our previous measurements of the Stokes peaks of bP at 369 cm$^{-1}$, 443 cm$^{-1}$, and 469.2 cm$^{-1}$ \cite{Tayari2015}, is likely originating from heating due to the elevated pump power required to acquire Raman spectra from gold encapsulated bP.\\

\begin{figure}[t]
	\begin{center}
		\includegraphics[width=.6\linewidth]{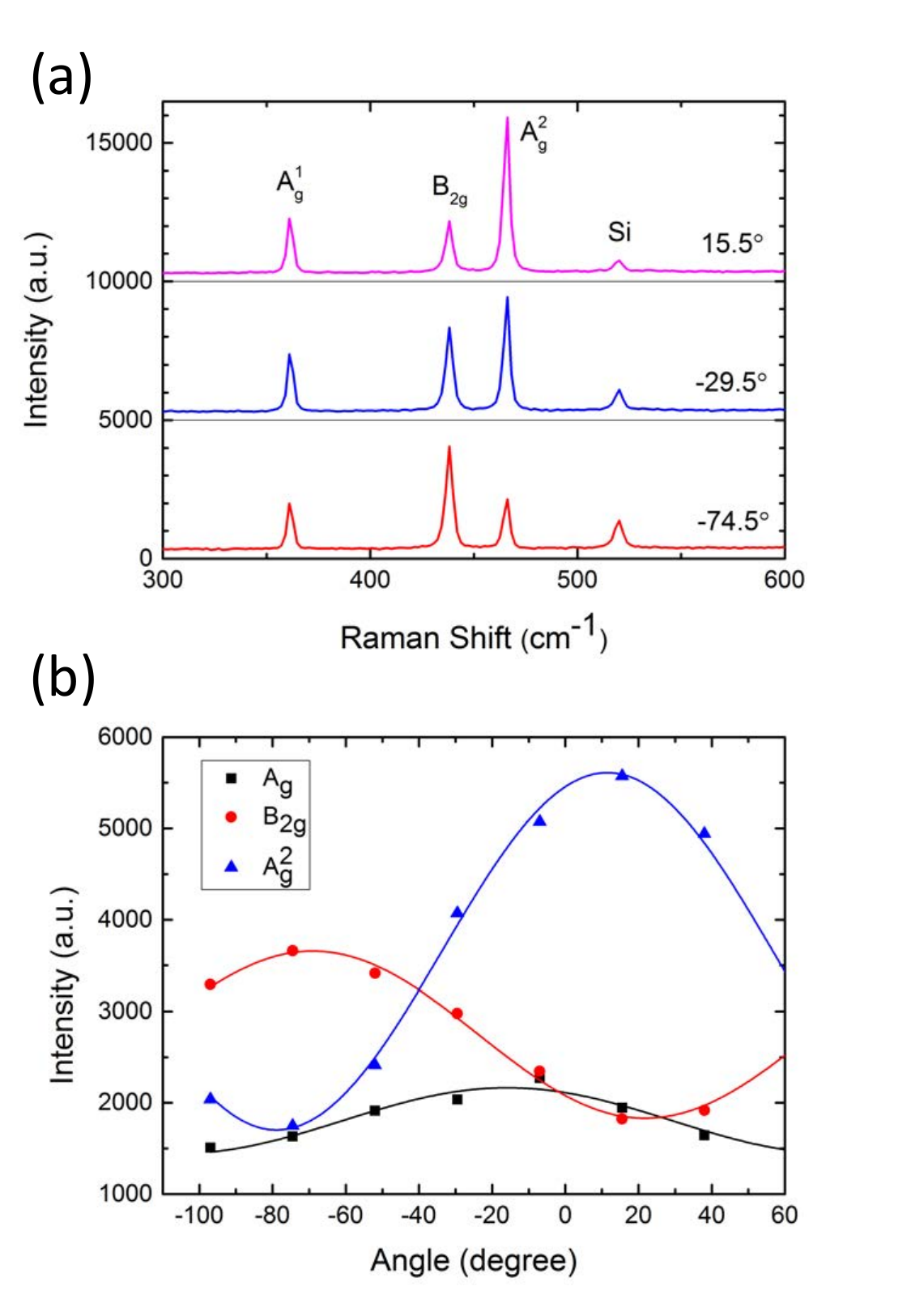}
		\caption{{\bf Polarized Raman measurements}.  a) Three Raman spectra taken under different polarization angles, 15.5$^\circ$, -29.5$^\circ$, and -74.5$^\circ$. b) Intensities of the three bP Raman peaks as a function of the polarization angle. The fits are guides to the eye. The laser spot size is $1~\mu m$ in diameter, and has a wavelength $\lambda= 532$~nm.}
		\label{fig:PolarizedRaman}
	\end{center}
\end{figure}

\begin{figure}[t]
	\begin{center}
		\includegraphics[width=1.\linewidth]{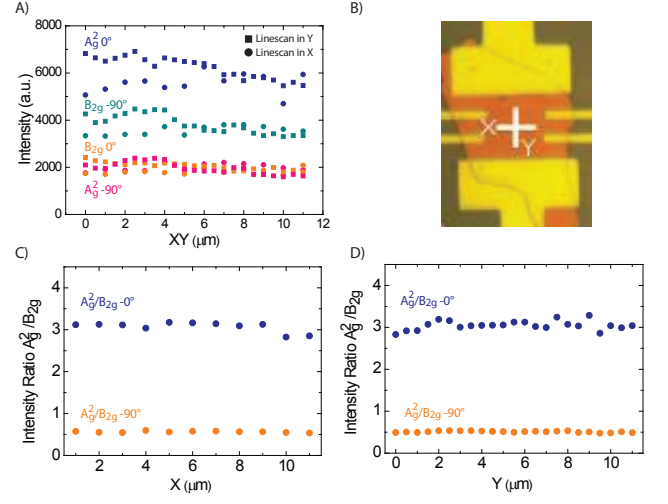} 
		\caption{{\bf Homogeneity of the sample verified by micro-Raman}. a) $A^2_g$ and $B_{2g}$ intensities for linescans in the $X$ and $Y$ directions and polarization angles of 0$^{\circ}$ and -90$^{\circ}$. For each linescan, the solid symbols refer to 0$^{\circ}$ polarization whereas the open symbols refer to -90$^{\circ}$ polarization. The blue color denotes the $B_{2g}$ peak whereas red is for  t$A^2_g$. b) Optical microscopy image of the device with bP flake and the XY directions indicated. The intensity ratio  $A^2_g / B_{2g}$ measured along  the $X$ and $Y$ direction is shown in panel c) and d), respectively. }
		\label{fig:RamanAngle}
	\end{center}
\end{figure}

\begin{figure}[t]
	\begin{center}
		\includegraphics[width=.8\linewidth]{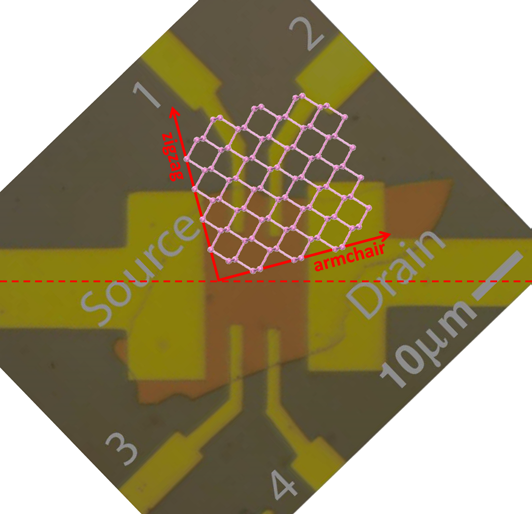} 
		\caption{{\bf Orientation of the bP crystal with respect to the source and drain}. From the polarized Raman data, an angle corresponding to  $\sim$15$^\circ$ was determined between the zigzag and the source-drain axis.  }
		\label{fig:OrientationCartoon}
	\end{center}
\end{figure}

\begin{figure}[t]
	\begin{center}
		\includegraphics[width=1.08\linewidth]{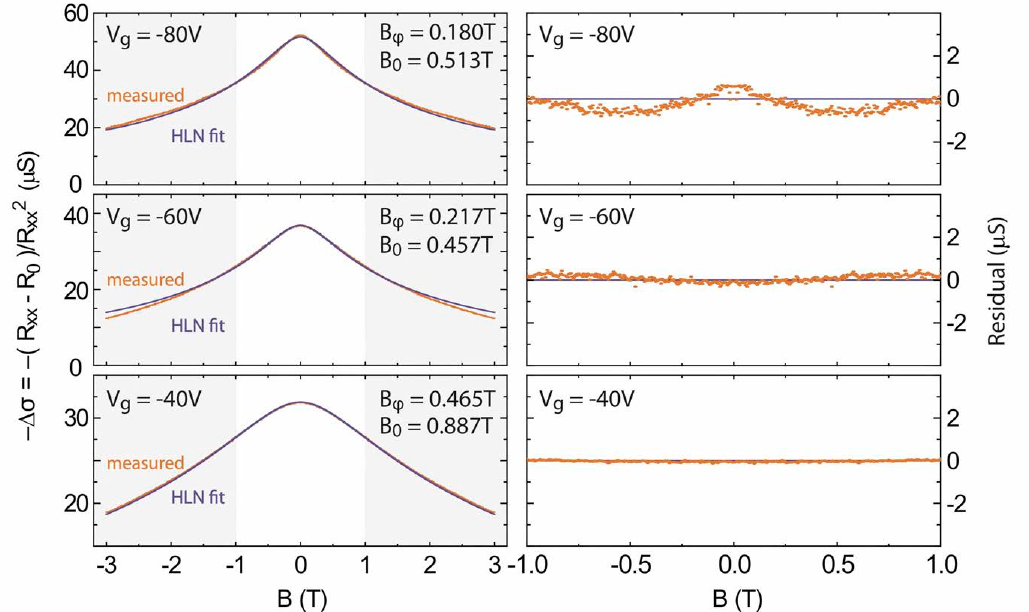}
		\caption{{\bf HLN weak localization fits at 10K}. The measured longitudinal conductance correction $-\Delta \sigma$ versus magnetic field $B$ and a best fit to a Hikami-Larkin-Nagaoka theory over the field range $-1~\mathrm{T}<B<1~\mathrm{T}$, where weak localization dominates, is shown for various gate voltages $V_g$. The extracted characteristic field $B_{0}$ for elastic and $B_{\phi}$ for inelastic (dephasing) scattering are indicated.}
		\label{fig:HLN10}
	\end{center}
\end{figure}

{\bf Polarized Raman Characterization}\\

Polarized Raman spectroscopy was performed using a Renishaw inVia system equipped with a 532 nm laser, a half-wavelength retarder (half-wave plate), and a motorized stage for 2D mapping of samples. A laser spot size of approximately $1 \mu m$ in diameter was used.  Figure \ref{fig:PolarizedRaman} a) shows three different spectra measured at different polarization angle with respect to the source-drain direction. At 0$^{\circ}$, the laser is polarized along the source-drain direction. Figure \ref{fig:PolarizedRaman} b) shows the variation of the Raman peak intensities with polarization angle. The maximum of the $A^2_g$ peak is known to be along the armchair direction \cite{Sugai1985,Wang2015}, and it was deduced that the bP  flake is oriented with an angle of  approximately 15$^{\circ}$ with respect to the source-drain direction. A photograph of the device is shown in Fig.\ref{fig:OrientationCartoon} showing the cystal orientation with respect to the source-drain axis of the current leads.  \\

Subsequently, micro Raman measurements  were performed to probe the homogeneity of the bP flake. In Fig.\ref{fig:RamanAngle} , the intensities of the $A^2_g$ and $B_{2g}$ peaks are shown as a function of the position on the sample for incident laser polarizations of $Y = 0^{\circ}$ ({\it i.e.} source-drain direction) and $X = -90^{\circ}$ ({\it i.e.} perpendicular to source-drain direction). The axis convention is shown in Fig.\ref{fig:RamanAngle} b). These two peaks were selected since their modulation with polarization angle was strongest. In panels c) and d) of Fig.\ref{fig:RamanAngle}, the $A^2_g/B_{2g}$ ratio is displayed for the two different polarization directions along the X and Y direction, respectively. The constant intensity ratio observed for a given polarization angle is strong indication that the bP flake is a single crystal.

{\bf Additional weak localization measurements and HLN fits}\\
For completeness, additional data taken at 10K are shown in Fig.\ref{fig:HLN10}  together with the HLN fits and residuals. The fitting procedure and the legend convention of the figure  is identical to that described in the main text.\\

\bibliography{Nick_supp}

\begin{thebibliography}{40}
\bibitem [{ICP()}]{ICP}%
  \BibitemOpen
  \href@noop {} {}\bibinfo {note} {We have performed an analysis of the bP
  crystal used in these experiments by Inductively Coupled Plasma-Mass
  Spectrometry (ICP-MS) which revealed a negligible level of magnetic and heavy
  elements, below 1 $\permil$. In addition, phosphorus being a light element,
  its spin orbit coupling is negligeably small.}\BibitemShut {Stop}%
\end{thebibliography}

\end{document}